\documentclass[twoside,11pt]{article}

\usepackage{jmlr2e}

\usepackage{longtable}
\usepackage{booktabs}
\usepackage[load-configurations=version-1]{siunitx} 
\usepackage{multirow}
\usepackage{multicol}
\usepackage{graphicx}
\usepackage{amsmath}

\newcommand{\dataset}{MelodySim}
\newcommand{\model}{MelodySim}

\begin{document}

\title{MelodySim: Measuring Melody-aware Music Similarity for Plagiarism Detection}

\author{\name Tongyu Lu\thanks{Equal contribution} 
\email{tongyu\_lu@mymail.sutd.edu.sg} \\
\addr Singapore University of Technology and Design
\AND
\name Charlotta-Marlena Geist\textsuperscript{*} 
\email{charlotta-marlena.geist@ovgu.de} \\
\addr Otto von Guericke University Magdeburg
\AND
\name Jan Melechovsky 
\email{jan\_melechovsky@mymail.sutd.edu.sg} \\
\addr Singapore University of Technology and Design
\AND
\name Abhinaba Roy 
\email{abhinaba\_roy@sutd.edu.sg} \\
\addr Singapore University of Technology and Design
\AND
\name Dorien Herremans 
\email{dorien\_herremans@sutd.edu.sg} \\
\addr Singapore University of Technology and Design
}


\maketitle

\begin{abstract}
We propose \textbf{\dataset{}}, a melody-aware music similarity model and dataset for plagiarism detection. First, we introduce a novel method to construct a dataset focused on melodic similarity. By augmenting Slakh2100, an existing MIDI dataset, we generate variations of each piece while preserving the melody through modifications such as note splitting, arpeggiation, minor track dropout, and re-instrumentation. A user study confirms that positive pairs indeed contain similar melodies, while other musical tracks are significantly changed. Second, we develop a segment-wise melodic-similarity detection model that uses a MERT encoder and applies a triplet neural network to capture melodic similarity. The resulting decision matrix highlights where plagiarism might occur. The experiments show that our model is able to outperform baseline models in detecting similar melodic fragments on the MelodySim test set.
\end{abstract}

\begin{keywords}
MelodySim, music similarity, plagiarism detection, Triplet neural network, dataset augmentation
\end{keywords}

\section{Introduction}
\label{sec:intro}
In recent years, generative music models have gained widespread popularity. Commercial models such as Suno\footnote{\url{https://suno.com}} and Udio\footnote{\url{https://udio.com}}, alongside open-source models like Mustango \citep{melechovsky2024mustango}, JAM \citep{liu2025jam} and MusicGen \citep{copet2023simple} have emerged in recent past. Symbolic music also has seen some text-to-MIDI generation models \citep{bhandari2025text2midi, roy2025text2midi}. This has raised the important questions regarding artist protection. This has sparked ongoing discussions and legal battles regarding how artists should be compensated for the use of their music as training data \citep{wei2024prevailing, herremans2025royalties}, exemplified by the Recording Industry Association of America (RIAA) lawsuit against Udio and Suno in June 2024 \citep{RIAA_2024_landmark_ai_cases}.
Moreover, the music generated by these models may also plagiarize the original training data. In this work, we offer one of the first efforts towards making tools to assist in melody-related plagiarism detection.  

Training generative models on copyrighted data (often improperly licensed) creates a strong possibility that the generated music plagiarizes the original training data. Diffusion models, in particular,  are prone to replicating their training data, as demonstrated by \cite{somepalli2023diffusion, carlini2023extracting} in image generation tasks. Artists have publicly highlighted instances where their work or style has been replicated by generative models \citep{Reed2023_AI_song_mimicking_Drake_Weeknd}.
The literature reveals that generative AI models are typically evaluated in terms of their ability to predict similarly to the input data (accuracy) rather than in terms of the originality of the generated outputs \citep{sturm2019artificial}. While no clear legal precedent currently addresses copyright issues regarding input data, we can examine potential plagiarism in the outputs of generative models. 

Identifying and confirming music plagiarism is inherently a complex task. \cite{huber2020armseligen} highlighted the necessity of considering each case individually. 
Automatic plagiarism detection tools could help speed up both the  identification of new plagiarism cases and the validation of expert opinions in past legal proceedings. Such tools could even be integrated directly into the music generation models to prevent plagiarized outputs. However, this task is not trivial, as there is no generally accepted, objective definition of music plagiarism. In analyzing 17 lawsuits, \cite{huber2020armseligen} observed that the melody was the primary consideration in determining plagiarism, followed by an `overall impression' of the music. This observation motivates the development of a melody-aware music similarity tool. However, existing works on melody similarity are limited to symbolic music (MIDI) \citep{karsdorp2019learning,park2022music,park2024quantitative}. To address real-world court cases and generated music, we develop an audio-based melody-aware similarity model in this work. This presents greater challenges than symbolic approaches due to overlapping audio signals and limited training data availability. 

This work makes several key contributions. First, we introduce MelodySim, a novel dataset consisting of 1,568 full-length instrumental (audio) songs, each with three variations, resulting in a total of 6,272 files. These variations incorporate subtle melodic changes, including changes in note durations, speed, and instrumentation, while remaining sufficiently similar to be considered potentially plagiarized. Second, we train a Triplet Neural Network with a MERT encoder \citep{li2023mert} on this dataset, optimizing it to minimize representation distances between matching segments while maximizing distances between non-matching ones. This results in melody-aware similarity embeddings, which are then used in a classification model to predict segment matches. Such metrics could prove valuable for copyright attribution mechanisms in generative models, as proposed by \citet{herremans2025royalties}.

The remainder of this paper is organized as follows. Section~\ref{sec:related} reviews related work, and Section~\ref{sec:dataset} describes the construction of the MelodySim dataset. We then present our melody-aware triplet neural network architecture and full-song plagiarism detection method. Finally, we report experimental results on both our test set and in-the-wild plagiarism cases before concluding with a discussion of limitations and future work.

\section{Related Work}
\label{sec:related}
In this section, we provide a brief overview of how music plagiarism has been defined in existing literature. We then discuss related work on music similarity detection models.

\subsection{What is plagiarism?}

When developing a plagiarism detection model, we must first ask: \emph{which musical elements constitute plagiarism?} Many popular songs share common features such as I–IV–V harmonic progressions or reused samples, so similarity in harmony or instrumentation alone is insufficient evidence for copyright infringement claims. This leaves other musical identities arising from melody, rhythm, harmony, timbre, and structure as potential sources of plagiarism.

No fixed rule set currently defines musical plagiarism. \cite{yesiler2021audio} showed that “likely the same” or “minor variation” pairs usually preserve melody and harmony, while dissimilar pairs diverge in melody. \cite{huber2020armseligen} analyzed 17 music plagiarism lawsuits and found that melody was consistently prioritized in plagiarism determinations, typically in conjunction with another parameter—most often rhythm. The authors noted that melody is the most debated aspect in legal disputes, second only to overall impression, which encompasses the interplay of various musical characteristics.

Based on these findings, we focus on melody as the primary cue of similarity, while still representing the broader musical context through MERT features~\citep{li2023mert}. To enable this, we created a controlled dataset that systematically modifies musical attributes at multiple levels while preserving most of the original melody, as described in Section~\ref{sec:dataset}.

\subsection{Automatic Music Similarity Detection}


\paragraph{Symbolic Domain.}
 Most existing work on \emph{melody} similarity domain has been done in the symbolic domain, often targeting melody retrieval or duplication detection rather than plagiarism. For example, \cite{karsdorp2019learning} trained recurrent networks on the Meertens Tune Collections~\citep{van2014meertens} to rank melodies by similarity, while \cite{yin2021good} proposed an originality score to measure overfitting in symbolic music generation. \cite{park2022music} represented MIDI excerpts as grayscale images to detect rhythmic and melodic plagiarism, but restricted the task to monophonic instrumental tracks. An important real-world resource is the Music Copyright Infringement Cases (MCIC) dataset~\citep{park2024quantitative} which contains music pairs from 116 actual court cases (66 denied, 32 infringed, 18 settled) in both MIDI and score formats.

\paragraph{Audio domain.}
Audio-based similarity methods are less common for plagiarism detection. Spectrogram- and fingerprint-based techniques~\citep{borkar2021music,lopez2022proposal} require large databases and degrade under noise or compression, as they capture general acoustic resemblance rather than specific musical structure. Dynamic Time Warping (DTW)~\citep{muller2007dynamic} and its variants have also been employed to align chroma or spectral features between recordings, enabling comparison despite tempo variations. However, such methods rely purely on low-level acoustic or tonal similarity and lack awareness of higher-level musical semantics such as phrasing or melodic intent. The MiRA tool~\citep{batlle2024towards} compares raw-audio embeddings for replication assessment, but its scope is limited to near-identical excerpts. \citet{kasif2023exploring} addressed sample reuse with a Siamese CNN trained on mel-spectrograms using a triplet loss, focusing on exact repetition rather than melodic variation. 

\paragraph{Cover-song and version identification.}
Related research in \emph{cover-song} or \emph{musical-version identification} aims to recognize different renditions of the same composition, typically using beat-synchronous chroma features and dynamic-time-warping alignment, or compact audio fingerprints for large-scale retrieval~\citep{yesiler2021audio}. While highly effective for matching recordings at the audio level, such methods primarily exploit low-level spectral or harmonic similarity and lack explicit modeling of higher-level musical semantics (e.g., note identity, melodic contour, structural motifs). MelodySim complements these approaches by evaluating whether audio embeddings can capture semantic, melody-aware similarity rather than purely acoustic resemblance.

\paragraph{Our contribution.}
We address the gap between symbolic and audio-domain research by introducing the first open, large-scale synthetic \emph{audio} dataset for melody similarity and plagiarism analysis. 
Each piece includes three music theory-informed melodic variations with fine-grained changes in pitch, rhythm, and articulation, while other instrumental tracks and timbre are substantially altered. These controlled yet subtle variations provide realistic test cases for assessing melody-aware similarity models trained directly on audio.

\section{MelodySim Dataset}
\label{sec:dataset}

\label{data_generation}

\begin{figure*}[h]
    \centering
    \includegraphics[width=1\linewidth]{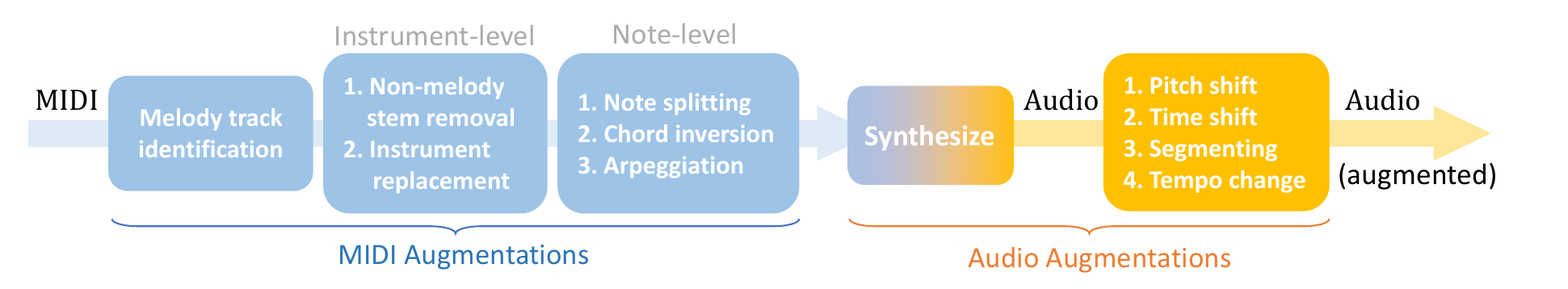}
    \caption{The proposed melody-aware augmentation pipeline used for constructing the \dataset{} dataset by augmenting Slakh MIDI.}
    \label{fig:augmentation_flowchart}
\end{figure*}

Although large-scale datasets exist in both audio \citep{bogdanov2019mtg, roy2025jamendomaxcaps} and MIDI domains \citep{melechovsky2024midicaps, raffel2016learning}, none provide paired examples with controlled melodic similarity and non-melodic variation required for training melody-aware plagiarism detection models. To address this gap, we develop the \textbf{\dataset{}} dataset, a novel audio dataset containing three variations per song through systematic MIDI and audio augmentations. These variations preserve melodic content (with minor changes for robustness) while modifying other aspects including track removal, instrument changes, chord inversions, tempo adjustments, and transposition, as illustrated in Figure~\ref{fig:augmentation_flowchart}. This approach enables us to capture melodic similarity across otherwise distinct songs, aligning with melody's central role in plagiarism determination \citep{huber2020armseligen}.

We use 1,568 MIDI files from the Slakh2100 dataset~\citep{manilow2019cutting} as the foundation for these augmentations. The following subsections detail our augmentation procedure. The resulting dataset comprises 6,272 full-length audio files, each original piece accompanied by three variations. Dataset and the augmentation code are publicly available\footnote{\url{https://github.com/AMAAI-Lab/MelodySim}}. 

\subsection{Step 1 - Melody track identification}
We first identify the melody track in each multi-track MIDI file using a gradient boosting classifier model \citep{zheng2019melody} that we trained on the CMU Computer Music Analysis Dataset\footnote{\url{https://www.cs.cmu.edu/music/data/melody-identification/}}. We refined this dataset by manually correcting a subset of incorrect labels. The refined dataset is available online\footnote{\url{https://huggingface.co/datasets/amaai-lab/melodySim}}.

Building on the work by \cite{zheng2019melody}, we make several improvements to the input features. First, we added track-level features, including polyphony rate and note activation density. Second, we augment the feature set with average features computed across all other tracks in the same MIDI file, providing contextual reference information. Through cross-validation, we selected a histogram-based gradient boosting model that achieved 97\% accuracy on the CMU validation split\footnote{\url{https://github.com/lucainiaoge/midi-melody-identification}}. Manual inspection confirmed strong generalization to Slakh2100.

\subsection{Step 2 - MIDI-level augmentations}

With the melody tracks identified in Step 1, we apply a series of MIDI augmentations at both the instrument and note levels. 

\textbf{Instrument replacement:} We group the MIDI instrument indices (1-128) into ensembles ( e.g., pianos, guitars, high/low-register strings), and then reassign the track instruments using the following rules: 1. with probability 0.2, retain the instrument as it is; 2. if not, then with probability 0.7, change the instrument to another one in its ensemble (e.g., replacing piano with e-piano); 3. otherwise replace the instrument with another one in a different ensemble with similar pitch register; 4. ensure coupled tracks (e.g., piano tracks) to be applied with the same replacement policy; 5. avoid different instrument tracks being replaced into the same instrument.

\textbf{Track removal:} 1. with a probability drawn from a uniform distribution of [0.1, 0.5], for each track, mute the track; 2. with a probability of 0.5, mute the percussion track; 3. never mute the melody tracks (identified), bass tracks or other important tracks (vocals, piano or guitar accompaniment).

\textbf{Note splitting:} With a probability $p_{note,n}$, split the current note of typical duration (whole notes, half notes, quarter notes) into two of half the original duration. $p_{note,n}$ is drawn from a uniform distribution of [0.3, 0.85] for each track $n$.

\textbf{Chord inversion:} For each track, detect block chords (concurrently played notes) consisting of 3 or 4 notes. For each such chord, with a probability $p_{chinv,n}$, shift the top notes an octave down or the bottom notes an octave up. $p_{chinv,n}$ is drawn from a uniform distribution of [0.3, 0.85] for each track $n$.

\textbf{Chord arpeggiation:} For each track, detect block chords that are in regular durations (1x/2x/3x/4x quarter note). With a probability $p_{charg,n}$, split the chord into an arpeggio (consisting of equally-placed chord notes) with the same total duration as the original chord. $p_{charg,n}$ is drawn from a uniform distribution of [0.3, 0.85] for each track $n$.

\subsection{Step 3 - Audio-level augmentations}

After augmenting the MIDI files, the resulting audio files are obtained by synthesizing with the Musyng soundfont\footnote{available at \url{https://musical-artifacts.com}}. Then, a set of audio augmentations (Figure~\ref{fig:augmentation_flowchart}) is applied to further diversify the different versions, in particular:

\textbf{Pitch shift:} The audio track is pitch-shifted by a random integer of semitones in the range of [-4, 4].

\textbf{Time shift:} The whole track is shifted by a random time from a range of [-3, 3] seconds. This time shift is used when matching the positive pairs later on.

\textbf{Tempo change:} The audio track's tempo is altered by a random factor in the range of [0.9, 1.1].

While these transformations could have been applied at the MIDI level, implementing them in the audio domain enables future extension to real recordings where MIDI representations are unavailable, and allows experimentation with soundfont variations and acoustic effects. The resulting audio files are then cut into 10-second-long segments, each saved with their representative track name, version index, and segment index.

\section{Music similarity model}

Using the newly created \dataset{} dataset, we train a triplet neural network model \citep{hoffer2015deep} that enables the extraction of melody-sensitive embeddings of music audio, and the computation of the distance or similarity between these embeddings.

\subsection{Triplet dataset}
To train the model, we reformulate the \dataset{} dataset into triplets, each consisting of an anchor sample, a positive sample similar to the anchor, and a negative sample dissimilar to the anchor. The positive pairs are constructed by combining time-aligned segments from the original and augmented versions of the same track, ensuring that they share the same melody but differ in other aspects such as texture, tempo, or instrumentation. The negative pairs are formed by sampling inter-song segments from different tracks, which differ in both melody and other characteristics. For example, a triplet $(\text{anchor}, \text{positive}, \text{negative})$ may take the form:

\begin{align*}
\text{anchor} &= \text{Track}{00125}/\text{version}0/\text{segment}{02}, \\
\text{positive} &= \text{Track}{00125}/\text{original}/\text{segment}{02}, \\
\text{negative} &= \text{Track}{00007}/\text{version}2/\text{segment}{12}.
\end{align*}

\subsection{Triplet Neural Network}

\begin{figure*}[ht]
    \includegraphics[width=1\linewidth]{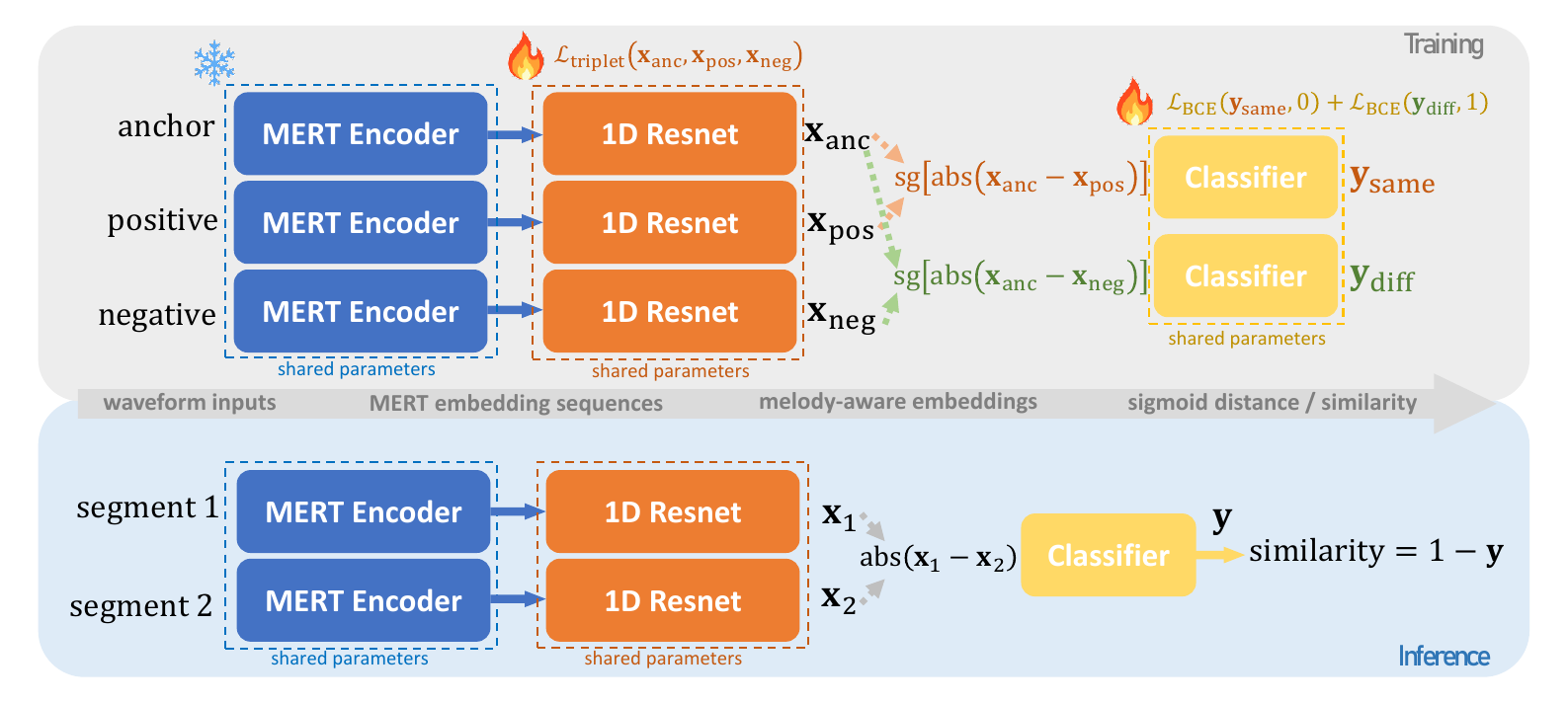}
    \caption{The proposed architecture for training and inference. $\mathrm{sg}[\cdot]$ means ``stop gradient" and $\mathrm{abs(\cdot)}$ notates element-wise absolute function.}
    \label{fig:model_architecture}
\end{figure*}

As illustrated in Figure~\ref{fig:model_architecture}, the proposed music similarity model is a triplet neural network (TNN) composed of a MERT encoder, a 1D ResNet backbone, and a classification head.

The audio input is first processed by a pretrained \verb|MERT-v1-95M| encoder~\citep{li2023mert}, selected for its balance between representational power and computational cost. While MERT captures both timbral and melodic information, this aligns with real-world plagiarism cases where overall musical impression matters alongside melodic similarity~\citep{huber2020armseligen}. The triplet loss fine-tuning adapts these rich representations toward melody-aware embeddings. To reduce memory usage, the extracted MERT features are averaged over time with a moving window (size=10, stride=10), and only four hidden layers (specifically, the outputs of MERT transformer layers 3, 6, 9, and 12, denoted $h_3, h_6, h_9, h_{12}$) are retained. The MERT parameters remain frozen during training.

The resulting MERT embeddings are passed to a lightweight 1D convolutional ResNet that adapts the MERT representations to melody-aware similarity embeddings. Average pooling over time yields a fixed-length vector for each segment. Given a triplet (anchor, positive, negative), the resulting similarity embeddings $\mathbf{x}_{\rm anc}, \mathbf{x}_{\rm pos}, \mathbf{x}_{\rm neg}$ are optimized with the triplet loss:
\begin{equation}
\mathcal{L}_{\rm triplet} =
\max \big(d(\mathbf{x}_{\rm anc}, \mathbf{x}_{\rm pos}) - d(\mathbf{x}_{\rm anc}, \mathbf{x}_{\rm neg}) + \alpha, 0 \big)
\label{eq:triplet_loss}
\end{equation}
where $d(\cdot)$ denotes the Euclidean distance and $\alpha=1.0$ is the margin.

In addition, a binary classifier is used to distinguish between ``same'' and ``different'' pairs. It receives the absolute embedding difference as input, $\mathrm{abs}(\mathbf{x}_{\rm anc}-\mathbf{x}_{\rm pos})$ or $\mathrm{abs}(\mathbf{x}_{\rm anc}-\mathbf{x}_{\rm neg})$, and outputs a similarity score in $[0,1]$ via a sigmoid layer. The classifier is trained using Binary Cross Entropy (BCE) loss:
\begin{equation}
\mathcal{L}_{\rm BCE} = \log(1-\mathbf{y}_{\rm same}) + \log(\mathbf{y}_{\rm diff})
\label{eq:bce_loss}
\end{equation}
where gradients are stopped at the classifier input to prevent interference with the embedding space ( $\mathrm{sg}[\cdot]$ in Figure~\ref{fig:model_architecture}).  

During inference, the system operates as a Siamese network: both audio segments are encoded and compared via absolute difference, yielding a final similarity score in $[0,1]$.

\subsection{Plagiarism Identification}

While the TNN measures similarity between two 10-second segments, plagiarism detection must operate at the piece level. We therefore compute a similarity matrix
\[
\mathbf{S}_{ij} = f(\mathbf{w}_1^{(i)}, \mathbf{w}_2^{(j)}) \in [0,1]^{N_1 \times N_2}
\]
where $\mathbf{w}_1^{(i)}$ and $\mathbf{w}_2^{(j)}$ are corresponding audio windows. A binary decision matrix is obtained via thresholding:
\[
\mathbf{D} = u(\mathbf{S} - \gamma)
\]
The threshold $\gamma$ reflects the model's confidence. we performed 2-fold cross-validation to set this threshold. $u$ is the Relu function. 

The row and column totals of $\mathbf{D}$ represent how many segments in one piece align between the two pieces. We can set a proportional threshold for when consider two pieces as too similar melody-wise. For example, when the proportional threshold is set to 0.2, then if more than 20\% of segments in both directions exceed the similarity threshold, the pair is considered as too-similar and flagged as potentially plagiarized. Note that although effective within the \dataset{} dataset domain and the MCIC dataset~\citep{park2024quantitative}, this rule-based aggregation is heuristic and may not generalize to diverse musical styles. In future research, we aim to improve the aggregation of segment-level results.

\section{Experimental Setup}
\subsection{Test Datasets}
We train the similarity model on 95\% of \dataset{} and evaluate on the remaining 5\%. We train on a single NVIDIA V100 GPU for 7.3 hours with batch size 512 over 797 epochs. We sample anchors with corresponding positive and negative pairs, revisiting each track four times per epoch to enhance diversity. The test split includes 78 tracks. For each track, we construct seven positive pairs among all version combinations (including self-comparisons), resulting in 546 positive pairs and an equal number of negative pairs from different tracks to ensure class balance.

We also evaluate on the MCIC dataset~\citep{park2024quantitative}, to assess whether our model generalizes and is able to detect songs that have been labeled as potential plagiarism cases in past copyright disputes.

\subsection{Baseline Models}

Since no prior work addresses melody-aware similarity following our augmentation strategy, we implement a naive Dynamic Time Warping (DTW) baseline to ground our evaluation. We first compute chroma and pitch contour features using the pYIN library~\citep{mauch2014pyin} and Constant-Q Transform (CQT)~\citep{brown1991calculation}. We then calculate the average DTW distance between pairs (mean over track lengths) and apply an adaptive threshold to maximize F1-score.

\section{Results}

We evaluate our approach on both the \dataset{} dataset test set and the out-of-distribution MCIC dataset, reporting objective metrics and subjective listening study results.

\subsection{Objective Evaluation}

\paragraph{Performance on MelodySim.} We evaluate the trained model at both the song and segment level (see Tables\ref{tab:seg} and \ref{tab:song} respectively). At the \textbf{segment-level} (Table~\ref{tab:seg}), the model achieves strong F1 scores, though class imbalance affects interpretation. The confusion matrix (Table~\ref{tab:conf}, left) reveals high precision (1.00) for negative (different) segments but lower precision (0.44) for positive (similar) segments, indicating the model conservatively flags similarity. Future threshold tuning could better balance precision-recall tradeoffs. 

\begin{table}[h!]
\centering
\small
\caption{Segment-level results on the \dataset{} test split.}
\label{tab:seg}
\setlength{\tabcolsep}{6pt}
\begin{tabular}{@{}lccc@{}}
\toprule
\textbf{Class} & \textbf{Precision} & \textbf{Recall} & \textbf{F1}  \\
\midrule
Different & 1.00 & 0.95 & 0.97 \\
Similar  & 0.44 & 0.99 & 0.61  \\
\textit{Weighted avg} & 0.98 & 0.95 & 0.96 \\
\bottomrule
\end{tabular}
\label{tab:melodysim_results_combined}
\end{table}
\begin{table}[h!]
\small
\caption{Confusion matrices for the segment-level (left) and song-level (right) results on the \dataset{} test set.}
\label{tab:conf}
\centering
\setlength{\tabcolsep}{4pt}
\[
\begin{array}{@{}c@{\hspace{1.5em}}c@{}}
\textbf{Segment-level} & \textbf{Song-level} \\[2pt]

\begin{tabular}{@{}l|l|c|c@{}|}
\multicolumn{2}{c}{} & \multicolumn{2}{c}{\textbf{GT}} \\ 
\cline{3-4}
\multicolumn{2}{c|}{} & \textbf{P} & \textbf{N} \\ 
\cline{2-4}
\multirow{2}{*}{\textbf{Pred.}} & \textbf{P} & 13{,}555 & 17{,}563 \\ 
\cline{2-4}
 & \textbf{N} & 81 & 319{,}824 \\ 
 \cline{2-4}
\end{tabular}
& 
\begin{tabular}{@{}l|l|c|c@{}|}
\multicolumn{2}{c}{} & \multicolumn{2}{c}{\textbf{GT}} \\ 
\cline{3-4}
\multicolumn{2}{c|}{} & \textbf{P} & \textbf{N} \\ 
\cline{2-4}
\multirow{2}{*}{\textbf{Pred.}} & \textbf{P} & 545 & 22 \\ 
\cline{2-4}
 & \textbf{N} & 1 & 524 \\ 
 \cline{2-4}
\end{tabular}

\end{array}
\]
\end{table}

At the \textbf{song-level}, we aggregate segment predictions and achieve a weighted F1-score of 0.98. We determine the aggregation threshold with 20 fold-cross cross-validation on MCIC, finding that 40\% segment overlap at 0.99 confidence level optimally distinguishes similar pairs. 
Table~\ref{tab:song} shows that our model clearly outperforms the DTW-based baselines (best: 0.79 for DTW with Chroma), demonstrating that the learned representations captures melodic similarity, even given small note variations, and varying instrumentation and accompaniment. The confusion matrix (Table~\ref{tab:conf}, right) shows only 1 false negative and 22 false positives among 1,092 test pairs, confirming robust song-level discrimination within the \dataset{} domain.

\begin{table}[h!]
\centering
\small
\caption{Song-level results on the \dataset{} test split.}
\label{tab:song}
\setlength{\tabcolsep}{5pt}
\begin{tabular}{@{}lcccc@{}}
\toprule
\textbf{Model} & \textbf{Precision} & \textbf{Recall} & \textbf{F1} & \textbf{AUC} \\
\midrule
\multicolumn{5}{l}{\textbf{\model{} (ours)}} \\
\quad Different & 1.00 & 0.96 & 0.98 & -- \\
\quad Similar   & 0.96 & 1.00 & 0.98 & -- \\
\quad \textit{Weighted avg} & 0.98 & 0.98 & \textbf{0.98} & -- \\
\addlinespace[1pt]
\midrule
DTW wih Chroma & 0.83 & 0.76 & 0.79 & 0.72 \\
DTW with Pitch  & 0.55 & 0.86 & 0.67 & 0.51 \\
DTW with CQT & 0.50 & 1.00 & 0.67 & 0.01 \\
\bottomrule
\end{tabular}
\label{tab:melodysim_results_combined}
\end{table}

\begin{table}[h!]
\centering
\small
\caption{Results on MCIC dataset. Our model uses thresholds from 20-fold cross-validation; DTW baselines use optimal thresholds per feature.}
\setlength{\tabcolsep}{4pt}
\begin{tabular}{@{}lcccccccc@{}}
\toprule
Model & TP & FP & TN & FN & Precision & Recall & F1 & Accuracy \\
\midrule
DTW with Chroma & 83 & 41 & 75 & 33 & 0.67 & 0.72 & 0.69 & 0.69 \\
DTW with Pitch & 105 & 86 & 30 & 11 & 0.55 & 0.91 & 0.68 & 0.58 \\
DTW with CQT & 94 & 51 & 65 & 22 & 0.65 & 0.81 & 0.72 & 0.71 \\
\midrule
\textbf{\model{} (overall)} & 101 & 61 & 55 & 15 & 0.62 & 0.87 & \textbf{0.73} & 0.67 \\
\textbf{\model{} (2-fold CV)} & 103 & 77 & 39 & 13 & 0.57 & 0.89 & 0.70 & 0.61 \\
\bottomrule
\end{tabular}
\label{tab:mcic_test}
\end{table}

\begin{table}[h!]
\centering
\small
\caption{Listening study results (mean opinion score $\pm$ 95\% CI).}
\setlength{\tabcolsep}{8pt}
\begin{tabular}{@{}lcc@{}}
\toprule
Aspect & Positive pairs & Negative pairs \\ 
\midrule
Overall similarity & $4.23\pm0.80$ & $2.00\pm0.68$\\
Melodic similarity & $4.53\pm0.84$ & $1.90\pm0.90$\\
Non-melodic similarity & $3.94\pm0.53$ & $2.27\pm0.22$\\
\bottomrule
\end{tabular}
\label{tab:survey}
\end{table}

\paragraph{Generalization to MCIC.}
We evaluate our model on the MCIC dataset, as shown in Table~\ref{tab:mcic_test}. The results show that the \model{} model is able to generalize moderately to other datasets, achieving an F1-score of 0.73, with most errors due to false positives. While the recall drops relative to results on the \dataset{} dataset, the result does indicate that the learned melody-aware representation is able to capture relevant musical similarities in a real-world plagiarism scenarios. This is encouraging, and motivates future research whereby we aim to expand the training set significantly to include more real-live audio (from more diverse distributions) as well as more melody augmentations. These modifications would make the model less prone to overfitting. 

\subsection{Subjective evaluation}
To validate dataset quality, we conducted a listening study with 12 participants who rated 12 randomly selected song pairs on overall, melodic, and non-melodic similarity using a 7-point Likert scale~\citep{joshi2015likert}. Table~\ref{tab:survey} shows that positive pairs received consistently higher ratings, confirming that our augmentations primarily alter non-melodic aspects while preserving melodic content.

\subsection{Discussion and future work}

\begin{figure}[h]
    \centering
    \includegraphics[width=0.7\linewidth]{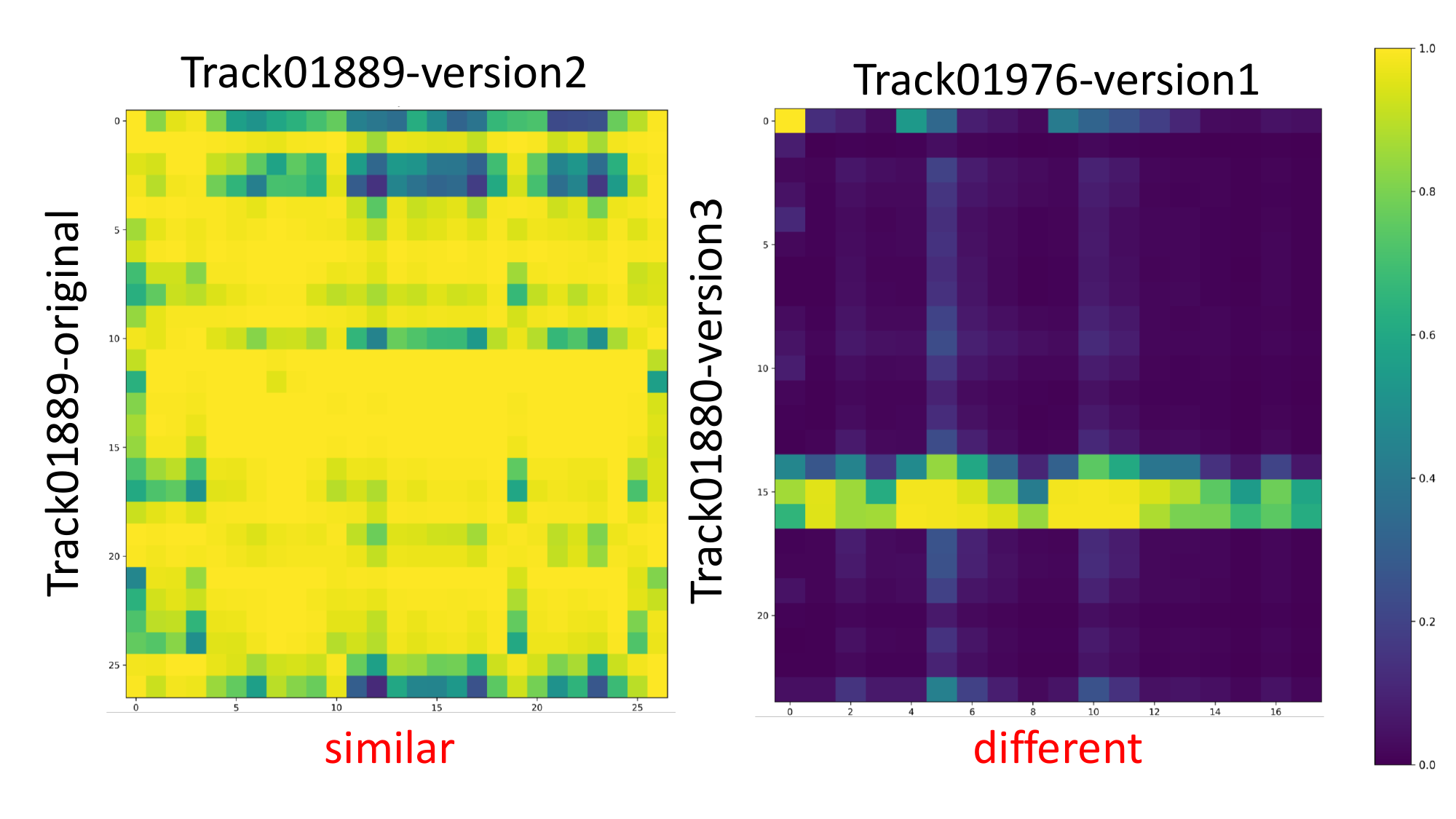}
    \caption{Similarity matrices for positive (left) and negative (right) pairs from the test set.}  
    \label{fig:similarity_matrix}
\end{figure}

Figure~\ref{fig:similarity_matrix} visualizes similarity matrices for two melodically similar (left) and dissimilar song pairs (right). Each pixel represents the predicted similarity between corresponding 10-second segments. 
The positive pair shows strong diagonal activation with broad surrounding similarity, while the negative pair exhibits only sparse, isolated matches. Such visualizations could aid court proceedings or help composers avoid unintentional similarity with existing works.

Our model and dataset represents the first step towards building an audio-based melody-aware semantic embedding for plagiarism detection. We believe that such embeddings could offer technical solutions for artist protection to the music industry in the generative AI era. For instance, such an embedding model might be used to attribute copyright based on training data similarity, flagging potential infringement cases, or guiding artists to avoid unintentional melodic overlap.

Several avenues for improvement remain. First, melody identification is challenging: tracks may contain multiple melodic voices or instruments carrying the melody intermittently, making rule-based extraction unreliable. Adaptive or learned melody extraction could address these limitations. Second, the performance gap between \dataset{} (0.98 F1) and MCIC (0.73 F1) reflects the controlled nature of synthetic augmentations. Training on more diverse data—including real plagiarism pairs, intra-song variations (verse vs. chorus), vocals, and broader musical styles—would improve robustness and generalization to authentic copyright scenarios.

\section{Conclusion}

We introduce \dataset{}, an open-source dataset and model for melody-aware music similarity and plagiarism detection. The dataset comprises original and augmented versions of tracks that maintain comparable melodies while varying in instrumentation and texture, verified through a listening study. Our MERT-based \model{} model, trained with a triplet loss, effectively detects melodic resemblance directly from audio, outperforming DTW-based baselines. Future directions include refining melody preservation in augmentation, incorporating intra-song variation, and improving the model generalization to authentic plagiarism scenarios.

\section{Acknowledgment}
This work has received support from SUTD’s Kickstart Initiative under grant number SKI 2021 04 06 and MOE under grant number MOE-T2EP20124-0014.
We acknowledge the use of ChatGPT for grammar improvements.

\bibliography{references}

\end{document}